\documentclass[copyright,creativecommons]{eptcs}
\providecommand{\event}{GandALF 2011} % Name of the event you are submitting to
\usepackage{breakurl}             % Not needed if you use pdflatex only.

\usepackage{latexsym}
\usepackage{setspace}
\usepackage{cancel}

\usepackage{graphicx}
\usepackage{appendix}
\usepackage{amssymb}
\usepackage{dsfont}
\usepackage{amsmath}
\usepackage{amsthm}
\usepackage{verbatim}
\usepackage{chngpage}
\usepackage{color}

\usepackage[noamssymb]{dlfltxbcodetips}

\newtheorem{theorem}{Theorem}
\newtheorem{definition}[theorem]{Definition}
\newtheorem{proposition}[theorem]{Proposition}
\newtheorem{corollary}[theorem]{Corollary}
\newtheorem{example}[theorem]{Example}
\newtheorem{lemma}[theorem]{Lemma}
\newtheorem{remark}[theorem]{Remark}

\def\arr#1{\stackrel{#1}{\longrightarrow}}

\def\arrud#1#2{\mathop{\longrightarrow}\limits^{#1}_{#2}}

\def\Dd{{\mathcal{D} }}

\def\data{{\mathds{D}}}

\def\Aa{{\mathcal{A} }}

\def\Bb{{\mathcal{B} }}

\def\Cc{{\mathcal{C} }}

\def\Ss{{\mathcal{S} }}

\def\Nn{\mathbb{N}}

\def\Ll{\mathcal{L}}

\def\Ii{\mathcal{I}}

\title{A Decidable Extension of Data Automata\footnote{This work was done while the author was a postdoc at LaBRI, Universit\'{e} Bordeaux 1, France.}}
\author{
Zhilin Wu
\institute{State Key Laboratory of Computer Science,\\
Institute of Software, Chinese Academy of Sciences\\
Beijing, China}
\email{wuzl@ios.ac.cn}
}
\def\titlerunning{A decidable extension of data automata}
\def\authorrunning{Zhilin Wu}
\begin{document}
\maketitle

\begin{abstract}
Data automata on data words is a decidable model proposed by Boja\'{n}czyk et al. in 2006. Class automata, introduced recently by Boja\'{n}czyk and Lasota, is an extension of data automata which unifies different automata models on data words. The nonemptiness of class automata is undecidable, since class automata can simulate two-counter machines. In this paper, a decidable model called class automata with priority class condition, which restricts class automata but strictly extends data automata, is proposed. The decidability of this model is obtained by establishing a correspondence with priority multicounter automata. This correspondence also completes the picture of the links between various class conditions of class automata and various models of counter machines. Moreover, this model is applied to extend a decidability result of Alur, Cern\'{y} and Weinstein on the algorithmic analysis of array-accessing programs.
\end{abstract}

%\vspace*{-3mm}
\section{Introduction}
%\vspace*{-2mm}

With the momentums from the XML document processing and the statical analysis and verification of programs, formalisms over infinite alphabets are becoming a research focus of theoretical computer science (c.f. \cite{Seg06} for a survey).

The infinite alphabet means $\Sigma \times \mathds{D}$, with $\Sigma$ a finite tag set and $\mathds{D}$ an infinite data domain. Words and trees with the labels of nodes from the infinite alphabet $\Sigma \times \mathds{D}$ are called data words and data trees. Formally, a data word is  a pair $(w,\pi)$, with $w$ denoting the sequence of tags and $\pi$ denoting the corresponding sequence of data values. Data trees can be defined similarly.

Among various models of logic and automata over infinite alphabets that have been proposed, data automata were introduced by Boja\'{n}czyk et al. in 2006 to prove the decidability of two-variable logic on data words (\cite{BMS+06}). 

A data automaton $\Dd$ consists of two parts, a nondeterministic letter-to-letter transducer $\Aa: \Sigma^\ast \rightarrow \Gamma^\ast$, and a class condition which is a finite automaton $\Bb$ with the alphabet $\Gamma$.  $\Dd$ accepts a data word $(w,\pi)$ iff from $w$, $\Aa$ is able to produce a $\Gamma$-string $w^\prime$ such that, 
\begin{quote}
for each class $X$ of $(w,\pi)$ (a class of a data word is a maximal set of positions with the same data value), $\Bb$ has an accepting run over $w^\prime \vert_X$ (the restriction of $w^\prime$ to the positions in $X$).   
\end{quote}

Several extensions of data automata have appeared in the literature. 

Extended data automata, was proposed by Alur, Cern\'{y} and Weinstein in 2009, in order to analyze the array-accessing programs (\cite{ACW09}). Extended data automata extend data automata by the class condition, which is now a finite automaton $\Bb$ with the alphabet $\Gamma \cup \{0\}$. $\Dd$ accepts a data word $(w,\pi)$ iff from $w$, $\Aa$ is able to produce a $\Gamma$-string $w^\prime$ such that, 
\begin{quote}
for each class $X$ of $(w,\pi)$, $\Bb$ has an accepting run over $w^\prime \oplus X$, where $w^\prime \oplus X$ is the string in $(\Gamma \cup \{0\})^\ast$ obtained from $w^\prime$ by replacing each letter $w^\prime_i$ such that $i \not \in X$ by $0$ (note that $w^\prime \oplus X$ has the same length as $w^\prime$). 
\end{quote}

However, as shown in \cite{ACW09}, it turns out that extended data automata are expressively equivalent to data automata, thus they are a syntactic extension, but not a semantic extension of data automata.

Another extension of data automata, class automata, was proposed by Boja\'{n}czyk and Lasota in 2010 to capture the full XPath, including forward and backward modalities and all types of data tests (\cite{BL10}).

Class automata generalize both data automata and extended data automata by the class condition, which is now a finite automaton $\Bb$ with the alphabet $\Gamma \times \{0,1\}$. $\Dd$ accepts a data word $(w,\pi)$ iff from $w$, $\Aa$ is able to produce a $\Gamma$-string $w^\prime$ such that, 
\begin{quote}
for each class $X$ of $(w,\pi)$, $\Bb$ has an accepting run over $w^\prime \otimes X$, where $w^\prime \otimes X$ is the string in $(\Gamma \times \{0,1\})^\ast$ obtained from $w^\prime$ by replacing each letter $w^\prime_i$ by $(w^\prime_i, 1)$ if $i \in X$, and by $(w^\prime_i, 0)$ otherwise. 
\end{quote}

In \cite{BL10}, Boja\'{n}czyk and Lasota also defined various class conditions of class automata and established their correspondences with different models of counter machines, including multicounter machines with or without zero tests, counter machines with increasing errors, and Presburger automata.  

Besides the models of counter machines considered in \cite{BL10}, there is still another type of counter machines, called priority multicounter automata, proposed by Reinhardt in his Habilitation thesis (\cite{Rei05}), where he showed that the nonemptiness of priority multicounter automata is decidable. Priority multicounter automata were also used by Bj\"{o}rklund and Bojanczyk to prove the decidability of two-variable first order logic over data trees of bounded depth (\cite{BB07}).

A priority multicounter automaton (PMA) is a multicounter automaton $M$ with the restricted zero tests: The $n$ counters in $M$ are ordered as $C_1, \dots, C_n$. $M$ can select an index $i \le n$, and test whether for each $j \le i$, $C_j=0$. 

In this paper, we propose a new type of class condition for class automata, called \emph{priority class condition}, and show its correspondence with priority multicounter automata, thus showing the decidability as well as completing  the picture of the links between class automata and counter machines established by Boja\'{n}czyk and Lasota.

The main idea of the priority class condition of class automata is roughly as follows: 
\begin{quote}
Let $\Dd=(\Aa,\Bb)$ be a class automaton such that the output alphabet of the transducer $\Aa$ is $\Gamma$. Then a priority class condition is obtained by putting an order (priority) over the letters $\gamma \in \Gamma$ and using this order to restrict the $(\gamma,0)$-transitions of $\Bb$.
\end{quote}

In this sense, a data automaton is a class automaton with priority class condition (PCA) in which all the $(\gamma,0)$-transitions are self-loops, while an extended data automaton is a PCA in which the different $\gamma$'s are non-distinguishable in $(\gamma,0)$-transitions.

With respect to the closure properties, we show that PCAs are closed under letter projection and union, but not under intersection nor  complementation. While data automata (and the expressively equivalent extended data automata) are closed under letter projection, union and intersection, it turns out that PCAs strictly extend data automata and still preserve the decidability. 

In addition, we demonstrate the usefulness of PCAs by applying them to generalize a decidability result of Alur, Cern\'{y} and Weinstein  on the analysis of array-accessing programs (\cite{ACW09}).

\medskip

This paper is organized as follows. In Section 2, some preliminaries are given. Then in Section 3, the concepts of $0$-priority finite automata and $0$-priority regular languages are introduced and PCA is defined. In Section 4, the correspondence between PCA and PMA is established. Section 5 discusses the application of PCAs to the algorithmic analysis of array-accessing programs. All the missing proofs can be found in the full version of this paper (\cite{Wu11}).

%\vspace*{-3mm}

\section{Preliminaries}

%\vspace*{-2mm}

In this paper, we fix a finite tag set $\Sigma$ and an infinite data domain $\data$, e.g. the set of natural numbers $\Nn$.

A \emph{word} $w$ over $\Sigma$ is a function from $[n]=\{1,\dots,n\}$ to $\Sigma$ for some $n \ge 1$. Suppose $w: [n] \rightarrow \Sigma$ is a word, then $|w|$ is used to denote the length of $w$, namely $n$. If in addition $X \subseteq [n]$, then $w |_X$ is used to denote the subword of $w$ restricted to the set of positions in $X$. A \emph{language} is a set of words.

A \emph{data word} is a pair $(w,\pi)$, where $w$ is a word in $\Sigma^\ast$ of length $n$ and $\pi: [n] \rightarrow \mathds{D}$. A \emph{class} of a data word $(w,\pi)$ (of length $n$) corresponding to a data value $d \in \data$ is a collection of all the positions $i \in \left[n\right]$ such that $\pi(i)=d$. For instance, the class of the data word $(a,0)(b,1)(c,0)$ corresponding to the data value $0$ is $\{1,3\}$. A \emph{data language} is a set of data words. Let $L$ be a data language, the language of words corresponding to $L$, denoted by $str(L)$, is $\{w \mid (w,\pi) \in L\}$.

%Let $prj: \Sigma \rightarrow \Sigma^\prime$, then the \emph{letter projection} of a data word $(w,\pi)$ under $prj$, %denoted by $prj((w,\pi))$, is $(prj(w_1),\pi_1)\dots (prj(w_{|w|}),\pi_{|w|})$, and the letter projection of a data language $L%$ under $prj$, denoted by $prj(L)$, is $\{prj((w,\pi)) \mid (w,\pi) \in L\}$. 

A \emph{data automaton} $\Dd$ consists of two parts, 
\begin{itemize}
\item a nondeterministic letter-to-letter transducer $\Aa: \Sigma^\ast \rightarrow \Gamma^\ast$,

\item and a class condition, which is a finite automaton $\Bb$ over the alphabet $\Gamma$.
\end{itemize}
A data automaton $\Dd=(\mathcal{A},\mathcal{B})$ accepts a data word $(w,\pi)$ iff from $w$, $\Aa$ is able to produce a string $w^\prime \in \Gamma^\ast$ (with the same length as $w$) such that for each class $X$ of $(w,\pi)$, $\Bb$ has an accepting run over $w^\prime \vert_X$.  The set of data words accepted by $\Dd$ is denoted by $\Ll(\Dd)$.

\emph{Class automata} $\Dd=(\Aa,\Bb)$ is an extension of data automata with the class condition changed into a finite automaton $\Bb$ over the alphabet $\Gamma \times \{0,1\}$. 

A class automaton $\Dd=(\Aa,\Bb)$ accepts a data word $(w,\pi)$ iff from $w$, $\Aa$ is able to produce a $\Gamma$-string $w^\prime$ such that for each class $X$ of $(w,\pi)$, $\Bb$ has an accepting run over $w^\prime \otimes X$, where $w^\prime \otimes X \in (\Gamma \times \{0,1\})^\ast$ is obtained from $w^\prime$ by replacing each letter $w^\prime_i$ by $(w^\prime_i, 1)$ if $i \in X$, and by $(w^\prime_i, 0)$ otherwise, e.g. if $w^\prime = a b c$ and $X=\{1,3\}$, then $w^\prime \otimes X =(a,1) (b,0) (c,1)$. The set of data words accepted by $\Dd$ is denoted by $\Ll(\Dd)$.

\medskip

A \emph{multicounter automaton} $\Cc$ is a hexa-tuple $(Q,\Sigma, k, \delta, q_0, F)$ such that
\begin{itemize}
\item $Q$ is a finite set of states,
\item $\Sigma$ is the finite alphabet,
\item $k$ is the number of counters,
\item $\delta \subseteq Q \times (\Sigma \cup \{\varepsilon\}) \times L \times Q$ is the set of transition relations over the instruction set $L=\{inc_i,dec_i,ifz_i \mid 1 \le i \le k\}$,
\item $q_0$ is the initial state,
\item  $F$ is the set of accepting states.
\end{itemize}

Let $\Cc=(Q,\Sigma, k, \delta, q_0, F)$ be a multicounter automaton. A \emph{configuration} of $\Cc$ is a state together with a list of counter values, namely, an element from $Q \times \Nn^k$. A configuration $(q^\prime,\overline{c^\prime})$ is said to be an \emph{immediate successor} of $(q,\bar{c})$ induced by a letter $\sigma \in \Sigma \cup \{\varepsilon\}$ and an instruction $l \in L$, denoted as $(q, \bar{c}) \xrightarrow{\sigma,l} (q^\prime,\overline{c^\prime})$, if $(q,\sigma,l, q^\prime) \in \delta$ and
\begin{itemize}
\item if $l=inc_i$, then $c^\prime_i = c_i+1$ and $c^\prime_j = c_j$ for $j \ne i$,
\item if $l=dec_i$, then $c_i > 0$, $c^\prime_i=c_i-1$, and $c^\prime_j = c_j$ for $j \ne i$,
\item if $l=ifz_i$, then $c_i=0$ and $c^\prime_j = c_j$ for each $j: 1 \le j \le k$.
\end{itemize}

A \emph{run} of $\Cc$ over a word $w$ is a nonempty sequence 
$(q_0, \overline{c_0}) \xrightarrow{\sigma_1,l_1} (q_1,\overline{c_1}) \xrightarrow{\sigma_2,l_2} \dots  \xrightarrow{\sigma_{n},l_{n}} (q_n,\overline{c_n})$ such that $w=\sigma_1 \dots \sigma_n$.
A run is \emph{accepting} if $q_n \in F$. $\Cc$ accepts a word $w$ if there is an accepting run of $\Cc$ over $w$.

A \emph{priority multicounter automaton} (abbreviated as PMA) is a multicounter automaton $\Cc$ with the following restricted zero tests: 
\begin{quote}
The $k$ counters in $\Cc$ are ordered as $C_1, \dots, C_k$. $\Cc$ can select some index $i \le k$, and test whether for each $j \le i$, the counter $C_j$ has value $0$. 
\end{quote}
Namely, a priority multicounter automaton is the same as a multicounter automaton, except that the instruction set $L$ is changed into $\{inc_i,dec_i,ifz_{\le i} \mid 1 \le i \le k\}$.

\begin{theorem}
[\cite{Rei05}]
 The nonemptiness of priority multicounter automata is  decidable.
\end{theorem}

%%%%%%%%%%%%%%%%%%%%%%%%%%%%%%%%%%%%%%%%%%%%%%%%%%%%%%%%%%
%%%%%%%%%%%%%%%%%%%%%%%%%%%%%%%%%%%%%%%%%%%%%%%%%%%%%%%%%%
%\vspace*{-4mm}
\section{Class automata with priority class condition}
%\vspace*{-2mm}

Intuitively, class automata with priority class condition are obtained from class automata by restricting the class condition to $0$-priority regular languages defined in the following.

\medskip

We first introduce several notations.

Let $\Bb=(Q, \Gamma \times \{0,1\}, \delta, q_0, F)$ be a \emph{deterministic complete} finite automaton over the alphabet $\Gamma \times \{0,1\}$. We use the notation $q \xrightarrow{(\gamma,b)} q^\prime$ to denote the fact that $\delta(q,(\gamma,b))=q^\prime$, where $b=0,1$, and $q \arr{\ast} q^\prime$ to denote the fact that $q^\prime$ is reachable from $q$ in the transition graph of $\Bb$. The transitions $q \arr{(\gamma,1)} q^\prime$ (resp. $q \arr{(\gamma,0)} q^\prime$) are called the one-transitions (resp. zero-transitions) of $\Bb$.

Let $G_0$ be the directed subgraph of the transition graph $(Q,\delta)$ obtained from $(Q,\delta)$ by restricting the set of arcs  to those labeled by letters from $\Gamma \times \{0\}$. Formally, $G_0=(Q, \delta \cap (Q \times (\Gamma \times \{0\}) \times Q))$. We use the notation $q \arrud{\ast}{0} q^\prime$ to denote the fact that $q^\prime$ is reachable from $q$ in $G_0$.

A state $q \in Q$ is called \emph{0-cyclic} if $q$ belongs to some nontrivial (containing at least one arc) strongly-connected component (SCC) $C$ of $G_0$.
Otherwise $q$ is called \emph{0-acyclic}.

For each $\gamma \in \Gamma$, let $G_{(\gamma,0)}$ be the directed subgraph of $(Q,\delta)$ obtained from $(Q,\delta)$ by restricting the set of arcs to those labeled by $(\gamma,0)$. Formally, $G_{(\gamma,0)}=(Q, \delta \cap (Q \times \{(\gamma,0)\} \times Q) )$. The out-degree of each vertex in $G_{(\gamma,0)}$ is exactly one, thus it has a simple structure: Each connected component of $G_{(\gamma,0)}$ consists of a unique cycle and a set of directed paths towards that cycle.

Let $\gamma \in \Gamma$. The cycles in $G_{(\gamma,0)}$ are called the \emph{$(\gamma,0)$-cycles} of $\Bb$. If a state $q$  belongs to some $(\gamma,0)$-cycle in $G_{(\gamma,0)}$, then $q$ is called a \emph{$(\gamma,0)$-cyclic} state, otherwise, it is called a \emph{$(\gamma,0)$-acyclic} state of $\Bb$. Note that $(\gamma,0)$-acyclic states may be $0$-cyclic.

\begin{example}
An example of the deterministic complete automaton $\Bb$ over the alphabet $\{a,b\} \times \{0,1\}$ is given in Figure \ref{fig-pca-example}(a). Its associated $G_0$ and $G_{(b,0)}$ are given in Figure \ref{fig-pca-example}(b) and Figure \ref{fig-pca-example}(c) respectively. The state $q_0$ and $q_2$ are both $0$-cyclic and $(b,0)$-cyclic, while $q_1$ is $0$-cyclic but $(b,0)$-acyclic, since $q_1$ belongs to a cycle in $G_0$ and does not belong to any cycle in $G_{(b,0)}$.
\begin{figure}[htbp]
\begin{center}
\includegraphics[width=0.9\textwidth]{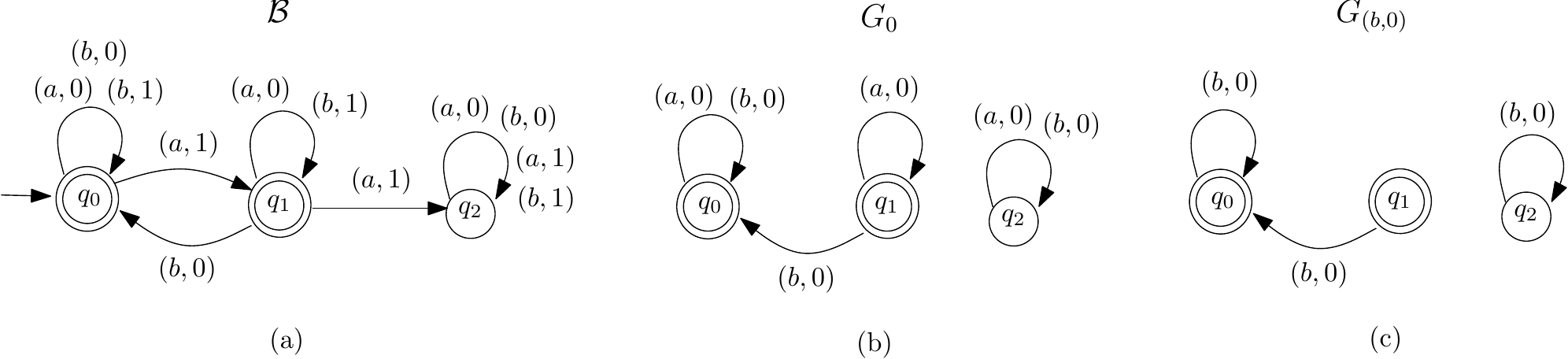}
\caption{Automaton $\Bb$, $G_0$ and $G_{(b,0)}$}
\label{fig-pca-example}
\end{center}
\end{figure}
\end{example}

%\vspace*{-1mm}
\begin{definition}[$((\gamma_1,0),(\gamma_2,0))$-pattern]
Let $\gamma_1,\gamma_2 \in \Gamma$. A \emph{$((\gamma_1,0),(\gamma_2,0))$-pattern} in $\Bb$ is a state-tuple $(q_1,q_2, q_3,q_4)$ such that  $q_1 \arr{(\gamma_1,0)} q_2 \arrud{\ast}{0} q_3 \arr{(\gamma_2,0)} q_4$, $q_1$ is $0$-cyclic, and $q_3$ is $(\gamma_2,0)$-acyclic.
\end{definition}

%\vspace*{-1mm}
\begin{example}
For the automaton $\Bb$ in Figure \ref{fig-pca-example}(a), because $q_1 \arr{(a,0)} q_1 \arrud{\ast}{0} q_1 \arr{(b,0)} q_0$, $q_1$ is $0$-cyclic and $(b,0)$-acyclic, it follows that $(q_1,q_1,q_1,q_0)$ is a $((a,0),(b,0))$-pattern in $\Bb$.
\end{example}

%\vspace*{-2mm}
\begin{definition}[$0$-priority finite automata and $0$-priority regular languages]
Let $\Bb$ be a finite automaton over the alphabet $\Gamma \times \{0,1\}$. Then $\Bb$ is called a $0$-priority finite automaton if $\Bb$ is a \emph{deterministic complete} automaton such that 
\begin{quote}
the letters in $\Gamma$ can be ordered as a sequence $\gamma_1 \gamma_2 \dots \gamma_k$ satisfying that there are no $((\gamma_i,0),(\gamma_j,0))$-patterns with $i \ge j$ in $\Bb$.
\end{quote}

A regular language $L \subseteq (\Gamma \times \{0,1\})^\ast$ is called a $0$-priority regular language if there is a $0$-priority finite automaton $\Bb$ over the alphabet $\Gamma \times \{0,1\}$ accepting $L$.
\end{definition}

Now we state several properties of $0$-priority finite automata and $0$-priority regular languages.

\begin{proposition}\label{prop-0-priority}
Let $\Bb=(Q, \Gamma \times \{0,1\}, \delta, q_0, F)$ be a deterministic complete finite automaton. Then $\Bb$ is a $0$-priority finite automaton iff $\Bb$ satisfies the following two conditions,

\begin{enumerate}
\item for any $\gamma \in \Gamma$, there are no $((\gamma,0),(\gamma,0))$-patterns in $\Bb$;

\item for any $\gamma_1,\gamma_2 \in \Gamma$ such that $\gamma_1 \ne \gamma_2$, if there is a $((\gamma_1,0),(\gamma_2,0))$-pattern in $\Bb$, then there do not exist $((\gamma_2,0),(\gamma_1,0))$-patterns in $\Bb$.
\end{enumerate}
\end{proposition}

%\vspace*{-2mm}
\begin{corollary}
Given a deterministic complete automaton $\Bb$ over the alphabet $\Gamma \times \{0,1\}$, it is decidable in polynomial time whether $\Bb$ is a $0$-priority finite automaton.
\end{corollary}

For each  nontrivial SCC, strongly-connected-component,  $C$ of $G_0$, let $L_{C}$ denote the set of labels $(\gamma,0)$ of the arcs belonging to $C$.

%Let $C_1, C_2$ be two distinct nontrivial strongly-connected components of $G_0$, $\gamma \in \Gamma$, if $q \arr{(\gamma,0)} q^\prime$ such that $q \in C_1$ and $q^\prime \in C_2$, then $C_1$ is said to be a \emph{$(\gamma,0)$-successor} of $C_2$, denoted as $C_1 \arr{(\gamma,0)} C_2$.

\begin{proposition}\label{prop-g-0}
If $\Bb$ is a $0$-priority finite automaton, then $G_0$ enjoys the following two properties. 
\begin{enumerate}
\item Suppose that $q_1 \arr{(\gamma,0)} q_2$ such that $q_1$ is $0$-cyclic, then $q_2$ is $(\gamma,0)$-cyclic. 
\item For each nontrivial SCC $C$ of $G_0$ and each $(\gamma,0) \in L_{C}$, every state in $C$ is $(\gamma,0)$-cyclic.
\end{enumerate}
\end{proposition}

From Proposition \ref{prop-g-0}, the following property can be easily deduced.
\begin{corollary}\label{cor-0-cyclic}
Let $\Bb$ be a $0$-priority finite automaton. If a state $q$ is reachable from some $0$-cyclic state in $\Bb$, then $q$ is $0$-cyclic as well.
\end{corollary}

In other words, the above corollary says that $0$-acyclic states cannot be reached from $0$-cyclic states in a $0$-priority finite automaton.

\begin{proposition}\label{prop-0-reg}
Let $L \subseteq (\Gamma \times \{0,1\})^\ast$ be a regular language. Then $L$ is a $0$-priority regular language iff the unique minimal deterministic complete finite automaton $\Bb$ accepting $L$ is a $0$-priority finite automaton.
\end{proposition}

\begin{definition}[Class automata with priority class condition, PCA]
A class automaton $(\Aa,\Bb)$ is said to have priority class condition, if the alphabet $\Gamma$ can be partitioned into $k$ ($k \ge 1$) disjoint subsets $\Gamma_1, \dots ,\Gamma_k$ such that $\Ll(\Bb)$ is a union of languages $L_1, \dots, L_k$ satisfying that $L_i \subseteq (\Gamma_i \times \{0,1\})^\ast$ is a $0$-priority regular language for each $i: 1 \le i \le k$.
\end{definition}

Intuitively, a class automaton $\Dd=(\Aa,\Bb)$ with priority class condition is a class automaton such that 
\begin{quote}
over a data word $(w,\pi)$, $\Aa$ nondeterministically chooses an index $i: 1 \le i \le k$, then produces a word $w^\prime \in \Gamma_i^\ast$, and verifies that each class string $w^\prime  \otimes X$ belongs to the $0$-priority regular language $L_i$.
\end{quote}

\begin{remark}
In the definition of PCAs, $\Ll(\Bb)$ is defined as a disjoint union of $0$-priority regular languages, instead of a single $0$-priority regular language. PCAs defined in this way can be shown to be closed under union (c.f. Proposition \ref{prop-closure}), while preserving the decidability (Theorem \ref{thm-main}). 
\end{remark}

%\vspace*{-1mm}

\begin{example}\label{exm-pca}
Let $\Cc$ be the class automaton $(\Aa,\Bb)$ such that $\Aa$ is the identity transducer and $\Bb$ is the automaton over the alphabet $\{a,b\} \times \{0,1\}$ in Figure \ref{fig-pca-example}(a). Then $\Cc$ accepts the data words satisfying the property ``between any two occurrences of the letter $a$ with the same data value, there is a letter $b$ with a different data value''. If $\{a,b\}$ is ordered as $ab$, then there are no $((a,0),(a,0))$-patterns, nor $((b,0),(a,0))$-patterns, nor $((b,0),(b,0))$-patterns, in $\Bb$. Thus $\Bb$ is a $0$-priority finite automaton under the ordering $ab$, so $\Cc$ is a PCA.
\end{example}

%\vspace*{-1mm}

\begin{remark}\label{rem-da-eda-ca}
Data automata can be seen as PCAs by adding self-loops $q \arr{(\gamma,0)} q$. Moreover, the extended data automata introduced in \cite{ACW09} can also be seen as a special case of PCA. In extended data automata, the class condition is a finite automaton $\Bb$ over the alphabet $\Gamma \cup \{0\}$, where the letters in $\Gamma$ are omitted in zero-transitions. Without loss of generality, $\Bb$ can be assumed to be deterministic and complete, then a deterministic complete finite automaton $\Bb^\prime$ over the alphabet $\Gamma \times \{0,1\}$ can be defined as follows: $q \arr{(\gamma,1)} q^\prime$ in $\Bb^\prime$ iff $q \arr{\gamma} q^\prime$ in $\Bb$, and $q \arr{(\gamma,0)} q^\prime$ in $\Bb^\prime$ iff $q \arr{0} q^\prime$ in $\Bb$. In the subgraph $G_0$ of $\Bb^\prime$, different letters $(\gamma,0)$ are non-distinguishable, so $G_0$ has the same structure as $G_{(\gamma,0)}$ for any $\gamma \in \Gamma$. Therefore, $\Bb^\prime$ is a $0$-priority finite automaton under any ordering of letters in $\Gamma$, and extended data automata can also be seen as PCAs.
\end{remark}

%\vspace*{-1mm}

\begin{proposition}\label{prop-closure}
The class of data languages accepted by PCAs are closed under letter projection and union, but not under intersection nor complementation.
\end{proposition}

The fact that PCAs are not closed under intersection is proved by contradiction:  If PCAs are closed under intersection, then PCAs are able to simulate two-counter machines, thus become undecidable, contradicting to Corollary \ref{cor-main} in the next section. 

Since data automata are closed under both union and intersection, it can be deduced that PCAs are strictly more expressive than data automata.

\begin{corollary}\label{cor-strict-more-express}
Class automata with priority class condition are strictly more expressive than data automata.
\end{corollary}

\begin{remark}
From Corollary \ref{cor-strict-more-express}, we know that there is a data language recognized by PCAs, but not by data automata. It would be nice if we could prove for instance that the data language in Example \ref{exm-pca}, namely, ``Between any two occurrences of the letter $a$ of the same data value, there is an occurrence of the letter $b$ with a different data value'', cannot be recognized by data automata. This is stated as an open problem in this paper.
\end{remark}

%%%%%%%%%%%%%%%%%%%%%%%%%%%%%%%%%%%%%%%%%%%%%%%%%
%%%%%%%%%%%%%%%%%%%%%%%%%%%%%%%%%%%%%%%%%%%%%%%%%

%\vspace*{-3mm}
\section{Correspondence between PCA and PMA}
%\vspace*{-2mm}

The aim of this section is to show that a correspondence between PCAs and PMAs can be established so that the decidability of the nonemptiness of PCAs follows from that of PMAs.

Let  $prj: \Sigma \rightarrow \Sigma^\prime \cup \{\varepsilon\}$, then the \emph{projection} of a data word $(w,\pi)$ under $prj$, denoted by $prj((w,\pi))$, is $prj(w_1)\dots prj(w_{|w|})$, and the projection of a data language $L$, denoted by $prj(L)$, is $\{prj((w,\pi)) \mid (w,\pi) \in L\}$. Note that the projection of a data language is a language, not a data language.

\begin{theorem}\label{thm-main}
The following two language classes are equivalent:
\begin{itemize}
\item projections of data languages accepted by PCAs,
\item languages accepted by PMAs.
\end{itemize}
\end{theorem}

\begin{corollary}\label{cor-main}
The nonemptiness of PCAs is decidble.
\end{corollary}

We prove Theorem \ref{thm-main} by showing the following two lemmas.

\begin{lemma}\label{lem-1to2}
For a PCA $\Dd$, a PMA $\Cc$ can be constructed such that $\Ll(\Cc)=str(\Ll(\Dd))$.
\end{lemma}
%\vspace*{-1mm}

From Lemma \ref{lem-1to2}, it follows that the first language class in Theorem \ref{thm-main} is included in the second one, since the class of languages accepted by PMAs is closed under mappings $prj: \Sigma_1 \rightarrow \Sigma_2 \cup \{\epsilon\}$. The next lemma says that the second language class in Theorem \ref{thm-main} is included in the first one.

\begin{lemma}\label{lem-2to1}
For a given PMA $\Cc$, a PCA $\Dd$ can be constructed such that $\Ll(\Cc)$ is a projection of $\Ll(\Dd)$.
\end{lemma}
%\vspace*{-1mm}

The rest of this section is devoted to the proof of the Lemma \ref{lem-1to2}.  The proof of Lemma \ref{lem-2to1} is omitted and can be found in the full version of this paper (\cite{Wu11}).

%\subsection{Proof of Lemma \ref{lem-1to2}}

The idea of the proof is to consider the abstract runs of class automata,  simulate them by multicounter automata, and illustrate that the simulation can be fulfilled by a priority multicounter automaton if the priority class condition is assumed. The proof is inspired by the proof of Theorem 2 in \cite{ACW09}.

%\vspace*{-2mm}
\subsection{From class automata to multicounter automata}\label{sec-ca-ma}

Let $\Dd=(\Aa, \Bb)$ be a class automaton, where $\Aa=(Q_g, \Sigma, \Gamma, \delta_g, q^g_0, F_g)$ and $\Bb=(Q_c,\Gamma \times \{0,1\}, \delta_c,q^c_0,F_c)$. Without loss of generality, we assume that $\Bb$ is deterministic and complete.

Given a data word $(w,\pi)$, let $\Ss(w,\pi)$ be the set of data values occurring in $(w,\pi)$, namely, $\Ss(w,\pi)=\{\pi_i \mid 1 \le i \le |w|\}$, and $(w,\pi)_{\le i}$ be the restriction of $(w,\pi)$ to the set of positions $\{1, \dots, i\}$ for each $i \le |w|$.

Intuitively, a run of $\Dd$ over a data word $(w,\pi)$ is a parallel running of the transducer $\Aa$ and the copies of the automaton $\Bb$ over $(w,\pi)$, with one copy for each data value occurring in $(w,\pi)$. A run of $\Dd$ over a data word $(w,\pi)$ can be seen as a sequence $(q^g_1,q^c_1, \gamma_1,R_1)(q^g_2, q^c_2,\gamma_2, R_2)\dots (q^g_{|w|}, q^c_{|w|}, \gamma_{|w|}, R_{|w|})$ such that 
\begin{itemize}
\item the sequence $(q^g_1, \gamma_1) \dots (q^g_{|w|}, \gamma_{|w|})$ corresponds to a run of the transducer $\Aa$,
\item $q^c_i$ records the state of a copy of $\Bb$ corresponding to a data value that has not been met until the position $i$, namely, a data value $d \not \in \Ss((w,\pi)_{\le i})$,

\item each time a new data value $\pi_i$ is met, $R_i(\pi(i))$ is set as $\delta_c(q^c_{i-1}, (\gamma_i,1))$, since $\pi(i)$ has not been met before and $q^c_{i-1}$ records the current state of $\Bb$ for the new data values.
\end{itemize}

Formally, A run of $\Dd$ over a data word $(w,\pi)$ is a sequence $(q^g_1,q^c_1, \gamma_1,R_1)\dots (q^g_{|w|}, q^c_{|w|}, \gamma_{|w|}, R_{|w|})$ satisfying the following conditions,
\begin{itemize}
%\item $(q^g_0, w_1, \gamma_1, q^g_1) \in \delta_g$, $\delta_c(q^c_0, (\gamma_1,0))=q^c_1$, 

\item for each $i: 1 \le  i \le |w|$, $(q^g_{i-1}, w_i, \gamma_i, q^g_{i}) \in \delta_g$, $\delta_c(q^c_{i-1}, (\gamma_i, 0))=q^c_{i}$ (where $q^g_0,q^c_0$ are the initial states of respectively $\Aa,\Bb$),

%\item $q_{|w|} \in F$, 

\item for each $i$, $R_i$ is a function from $\Ss((w,\pi)_{\le i})$ to $Q_c$, satisfying the following conditions,
\begin{itemize}
\item $R_1(\pi_1)=\delta_c(q^c_0, (\gamma_1,1))$,

\item for each $i: 1 < i \le |w|$, 

$R_{i}(\pi_i)=\delta_c(R_{i-1}(\pi_i), (\gamma_i, 1))$ if $\pi_i \in \Ss((w,\pi)_{\le i-1})$, otherwise $R_i(\pi_i)=\delta_c(q^c_{i-1}, (\gamma_i,1))$. 

For each $d \in \Ss((w,\pi)_{\le i-1})$ such that $d \ne \pi_i$, $R_{i}(d) = \delta_c(R_{i-1}(d),(\gamma_i,0))$.
\end{itemize}
\end{itemize}

A run $(q^g_1, q^c_1, \gamma_1,R_1)\dots (q^g_{|w|}, q^c_{|w|}, \gamma_{|w|}, R_{|w|})$ is \emph{successful} if $q^g_{|w|} \in F_g$ and $R_{|w|}(d) \in F_c$ for each $d \in \Ss(w,\pi)$.

The functions $R_1,\dots, R_{|w|}$ in a run of $\Dd$ on the data word $(w,\pi)$ can be abstracted into a sequence of functions $C_1, \dots, C_{|w|}$ such that each $C_i$ is a function $Q_c \rightarrow \mathds{N}$ satisfying that for each $q \in Q_c$, $C_i(q)$ is the number of data values $d \in \Ss((w,\pi)_{\le i})$ such that $R_i(d) = q$.

Intuitively, each $C_i$ is a tuple of counter values, with one counter for each state in $Q_c$. 
The sequence $C_1, \dots, C_n$ can be seen in a more abstract way, without directly  referring to the data values in $\Ss((w,\pi))$, as follows: 
%\vspace*{-1mm}
\begin{quote}
For each $1 < i \le |w|$, $C_i$ is obtained from $C_{i-1}$ by nondeterministically choosing one of the following two possibilities:
\begin{itemize}
\item either (corresponding to the situation $\pi_i \in \Ss((w,\pi)_{\le i-1})$)
\begin{itemize}
\item select some counter $q^\prime$ with non-zero value (i.e. $C_{i-1}(q^\prime) > 0$), decrement the counter $q^\prime$, 

\item then for each counter $q^{\prime\prime}$, the value of $q^{\prime\prime}$ is assigned as the sum of those of counters $p$ such that $\delta_c(p, (\gamma_i,0))=q^{\prime\prime}$, 

\item finally increment the counter $\delta_c(q^\prime,(\gamma_i,1))$.
\end{itemize}

\item or (corresponding to the situation $\pi_i \not \in \Ss((w,\pi)_{\le i-1})$)
\begin{itemize}
\item for each counter $q^{\prime\prime}$, the value of $q^{\prime\prime}$ is assigned the sum of those of counters $p$ such that $\delta_c(p, (\gamma_i,0))=q^{\prime\prime}$, 

\item increment the counter $\delta_c(q^c_{i-1},(\gamma_i,1))$.
\end{itemize}
\end{itemize}
\end{quote}

The sequence $(q^g_1,q^c_1, \gamma_1,C_1)(q^g_2, q^c_2,\gamma_2, C_2)\dots (q^g_{|w|}, q^c_{|w|}, \gamma_{|w|}, C_{|w|})$ is said to be an \emph{abstract run} of $\Dd$ over the data word $(w,\pi)$.  

%%%%%%%%%%%%%%%%%%%%%%%%%%%%%%%%%%%%%%%%%%%%%%%%%%%%%%%%%%%%%%%%%
%%%%%%%%%%%%%%%%%%%%%%%%%%%%%%%%%%%%%%%%%%%%%%%%%%%%%%%%%%%%%%%%%%

With such an abstract view of runs, $\Dd$ can be transformed into a multicounter automaton (with zero tests) $\Cc=(Q_a,\Sigma, k, \delta_a,q^a_0,F_a=\{q_{acc}\})$ as follows, 
\begin{itemize}
\item $Q_a$ includes $Q_g \times Q_c$ and some auxiliary states, e.g. for controlling the updates of the counter values.

\item $\Cc$ consists of $k=|Q_c|$ counters, one counter for each state in $Q_c$.

\item $q^a_0=(q^g_0,q^c_0)$.

\item Each $\gamma \in \Gamma$ induces a series of transition rules in $\delta_a$ as follows:

If 
%\vspace*{-2mm}
\begin{quote}
the current state of $\Cc$ is $(p^g,p^c)$, the read head is in a position labeled by $\sigma \in \Sigma$, and there are $q^g \in Q_g, q^c \in Q_c$ such that $(p^g,\sigma,\gamma,q^g) \in \delta_g$ and $\delta_c(p^c,(\gamma,0))=q^c$, 
\end{quote}
then 
%\vspace*{-2mm}
\begin{quote}
the state of $\Cc$ is changed into $(q^g,q^c)$, the counter values are updated in such a way to obtain $C_i$ from $C_{i-1}$ as above, and the read head is moved to the next position.
\end{quote}

\item Nondeterministically, $\Cc$ changes the state into a special state $q_s$ and repeats the following action: 
\begin{quote}
$\Cc$ arbitrarily chooses a non-zero counter $q \in F_c$, decrements $q$. Then it tests whether all the counters have zero value. If so, $\Cc$ changes the state into $q_{acc}$ and accepts.
\end{quote}
\end{itemize}

We now specify in detail how to update the counter values in $\Cc$, essentially, how to  perform the following updates: 
%\vspace*{-2mm}
\begin{quote}
For each counter $q^{\prime\prime}$ in $\Cc$, the value of $q^{\prime\prime}$ is assigned the sum of those of the counters $p$ such that $\delta_c(p, (\gamma,0))=q^{\prime\prime}$.
\end{quote}
%\vspace*{-2mm}
Recall that each connected component of $G_{(\gamma,0)}$ of $\Bb$ consists of a unique cycle $C$ and several paths towards $C$. Let $C=q_1 \dots q_r$, then for each $1 < i \le r$, the value of the counter $q_{i+1}$ is assigned as the sum of the value  of the counter $q_{i}$ and the values of the counters of its predecessors not in $C$, where $q_{r+1}=q_1$ by convention. Then the counter values can be updated as follows,
%\vspace*{-1mm}
\begin{enumerate}
\item the counters corresponding to the states in $C$ are first renamed\footnote{The idea of renaming is from \cite{ACW09}}: For each $i: 1 \le i \le r$, $q_i$ is renamed as $q_{i+1}$, where $q_{r+1}=q_1$ by convention. The renaming is remembered by the finite-state control of $\Cc$. With this renaming, the counter $q_{i+1}$ takes the value of the counter $q_i$ for each $i: 1 \le i \le r$.

\item then the values of the counters on the paths towards $C$ are updated in a backward way: For instance, let $p_1 \arr{(\gamma,0)} p_2 \arr{(\gamma,0)} p_3$ such that $p_3 \in C, p_1,p_2 \not \in C$, then the value of $p_2$ is first added into $p_3$, by decrementing $p_2$ and incrementing $p_3$ until the value of $p_2$ becomes zero; afterwards, the value of $p_1$ is added into $p_2$, and so on.
\end{enumerate}
%\vspace*{-1mm}

The above updates of counter values of $\Cc$ need (unrestricted) zero tests. In the following we will show that if $\Dd$ is a PCA, then these updates can be done with the restricted zero tests of PMAs, namely, testing zero for a prefix of counters as a whole, instead of a single counter. 

%\vspace*{-2mm}
\subsection{From PCA to PMA}
%\vspace*{-2mm}

We first assume that $(\Aa,\Bb)$ is a PCA such that $\Ll(\Bb)$ is a $0$-priority regular language, and $\Bb$ is a $0$-priority finite automaton. Later we will consider the more general case that $\Ll(\Bb)$ is a disjoint union of $0$-priority regular languages.

We first introduce some notations and prove a property of abstract runs of PCA.

Suppose that $\Gamma$ is ordered as $\gamma_1 \dots \gamma_l$ under which $\Bb$ is a $0$-priority finite automaton. 

Let $D_{scc}(G_0)$ be the strongly-connected-component directed graph of $G_0$ of $\Bb$, then $D_{scc}(G_0)$ is an acyclic directed graph. Let $\#_{scc}(G_0)$ denote the maximal length (number of arcs) of paths in $D_{scc}(G_0)$.

Similar to Lemma 1 in \cite{ACW09}, we can obtain the following lemma.
%
%\vspace*{-1mm}
\begin{lemma}\label{lem-0-cyclic}
Let $\Dd=(\Aa,\Bb)$ be  a PCA such that $\Bb$ is a $0$-priority finite automaton. Then any abstract run of $\Dd$ over a data word $(w,\pi)$, say $(q^g_1,q^c_1,\gamma_1,C_1)\dots(q^g_{|w|}, q^c_{|w|}, \gamma_{|w|},C_{|w|})$, enjoys the following property:

For each $i: 1 \le i \le |w|$, the sum of $C_i(q^\prime)$'s such that $q^\prime$ is $0$-acyclic is bounded by $\#_{scc}(G_0)$.
\end{lemma}
%\vspace*{-1mm}

By utilizing Lemma \ref{lem-0-cyclic}, we then demonstrate how the updates of the counter values of the multicounter automaton $\Cc$ obtained from $\Dd$ in Section \ref{sec-ca-ma} can be done with the restricted zero tests in PMAs.

We introduce some additional notations.

For each $i: 1 \le i \le l$, let $Acyc_i$ denote the set of \emph{$0$-cyclic} states $q \in Q_c$ such that $q$ is \emph{$(\gamma_i,0)$-acyclic}.

In addition, let $Acyc_{l+1}$ denote the set of $0$-cyclic states $q \not \in \bigcup \limits_{i: 1 \le i \le l} Acyc_i$ by convention.

%\vspace*{-2mm}
\begin{proposition}\label{prop-gamma-0-acyc}
Let $\Dd=(\Aa,\Bb)$ be  a PCA such that $\Bb$ is a $0$-priority finite automaton under the ordering $\gamma_1 \dots \gamma_l$. Then $Acyc_1, \dots, Acyc_{l+1}$ satisfy the following two properties:

\begin{enumerate}
\item $Acyc_i \subseteq Acyc_{i+1}$ for each $i<l$.

\item For each $i: 1 \le i \le l$, if $q \in Acyc_i$ and $q \arr{(\gamma_i,0)} q^\prime$, then $q^\prime \not \in Acyc_i$ and $q^\prime \in Acyc_j$ for some $j > i$. In particular, if $q \in Acycl_l$ and $q \arr{(\gamma_l,0)} q^\prime$, then $q^\prime \not \in Acyc_l$ and $q^\prime \in Acyc_{l+1}$.
\end{enumerate}
\end{proposition}

We are ready to show that if $\Dd$ is a PCA, then $\Cc$ can be turned into a PMA $\Cc_p=(Q_p,\Sigma, k, \delta_p, q^p_0,F_p)$.

From Lemma \ref{lem-0-cyclic}, if $\Dd$ is a PCA, then in the multicounter automaton $\Cc$, the sum of the values of the counters corresponding to the $0$-acyclic states of $\Bb$ are always bounded. Thus in $\Cc_p$, the counters corresponding to these $0$-acyclic states become \emph{virtual}, in the sense that the values of these counters are stored in the finite state control of $\Cc_p$,  and there are no \emph{real} counters in $\Cc_p$ corresponding to the $0$-acyclic states of $\Bb$.

The state set of $\Cc_p$ consists of the states $(p^g,p^c, \Ii_{Acyc})$ and some auxiliary states for updating the counter values, where $\mathcal{I}_{Acyc}$ is the information about the virtual counters corresponding to the $0$-acyclic states of $\Bb$. The counters of $\Cc_p$ correspond to the $0$-cyclic states of $\Bb$, with one counter for each $0$-cyclic state.

The counters (corresponding to the $0$-cyclic states of $\Bb$) of $\Cc_p$ are ordered according to the following order of $0$-cyclic states of $\Bb$,
\begin{center}
$Acyc_1 (Acyc_2 \setminus Acyc_1) \dots (Acyc_l \setminus Acyc_{l-1}) Acyc_{l+1},$
\end{center}
where an arbitrary ordering is given to the states within $Acyc_1$, $Acyc_{l+1}$, and each of $Acyc_{i+1} \setminus Acyc_{i}$ for $i: 1 \le i < l$.

Each $\gamma \in \Gamma$ induces a series of transition rules in $\delta_p$ specified in the following.

If the current state of $\Cc_p$ is $(p^g,p^c, \Ii_{Acyc})$, the read head is in some position labeled by $\sigma$, and there are $q^g \in Q_g, q^c \in Q_c$ such that $(p^g,\sigma,\gamma,q^g) \in \delta_g$ and $\delta_c(p^c,(\gamma,0))=q^c$, then the state of $\Cc_p$ is changed into $(q^g,q^c, \Ii^\prime_{Acyc})$. Now we illustrate how  the values of the real counters are updated and how the values of the virtual counters, i.e. $\Ii_{Acyc}$ in the finite state control of $\Cc_p$, is updated into 
$\Ii^\prime_{Acyc}$, by the following three steps.

\begin{enumerate}
\item  Either 
\begin{center}
the state $p^c_1=\delta_c(p^c, (\gamma,1))$ (a new data value is met) is stored in the finite state control of $\Cc_p$, 
\end{center}
or 
%\vspace*{-2mm}
\begin{quote}
some ($0$-acyclic or $0$-cyclic) state $q^\prime \in Q_c$ (an old value is met) is selected, the (virtual or real) counter  corresponding to $q^\prime$ is decremented, and the state $p^c_1=\delta_c(q^\prime, (\gamma,1))$ (the virtual or real counter corresponding to it should be incremented) is stored in the finite-state control of $\Cc_p$.
\end{quote}
%\vspace*{-1mm}
\item The values of the (virtual or real) counters are updated as follows.

Let $\gamma=\gamma_i$ for some $i: 1 \le i \le l$. 

The counters corresponding to the states in $Acyc_j \setminus Acyc_{j-1}$ for $j > i$, which are $(\gamma_i,0)$-cyclic in $\Bb$, are first updated by renaming, with the renaming stored in the finite state control of $\Cc_p$. Then for each counter $q \in Acyc_1$, the value of the counter $q$ is added to its $(\gamma_i,0)$-successor $q^\prime$, which is in $Acyc_j \setminus Acyc_i$ for some $j>i$ according to the fact that $q \in Acyc_1 \subseteq Acyc_i$, $q \arr{(\gamma_i,0)} q^\prime$ and Proposition \ref{prop-gamma-0-acyc}. Namely, the value of the counter $q$ is decremented and the value of $q^\prime$ is incremented until the value of the counter $q$ becomes zero. Afterwards, for each counter $q \in Acyc_2 \setminus Acyc_1$, the value of the counter $q$ is added to its $(\gamma_i,0)$-successor (which is also in $Acyc_j \setminus Acyc_i$ for some $j>i$), and so on, until all the counters corresponding to the states in $Acyc_i \setminus Acyc_{i-1}$ are updated. 

Note that during these updates of counter values, the zero-tests can be restricted to the zero-tests for a prefix of counters. The reason is that when updating the counter corresponding to a state $q \in Acyc_{j+1} \setminus Acyc_{j}$ for some $j < i$, the values of the counters corresponding to the states in $Acyc_1, \dots, Acyc_{j} \setminus Acyc_{j-1}$ are already zero. Therefore, testing zero for the counter $q$ is equal to testing zero for the counters before $q$ (including $q$) in the ordering. 

Then, $\mathcal{I}_{Acyc}$, i.e. the information about the values of the virtual counters, is updated into $\mathcal{I}^{\prime}_{Acyc}$ by following $G_0$, the zero-transitions of $\Bb$, and some real counters (corresponding to the $0$-cyclic states) should also  be incremented if they correspond to the $(\gamma_i,0)$-successors of some $0$-acyclic states in $\Bb$.

\item If $p^c_1$ is $0$-acyclic, then $\mathcal{I}^{\prime}_{Acyc}$ is further updated by incrementing the value of the virtual counter $p^c_1$, otherwise, the value of the real counter corresponding to the ($0$-cyclic) state $p^c_1$ is incremented.
\end{enumerate}

The definition of the $F_p$ of $\Cc_p$ is similar to $F_a$ of $\Cc$ in Section \ref{sec-ca-ma}.

Finally the read head is moved to the next position.

This finishes the description of $\Cc_p$.

\medskip

At last, we consider the general case that $\Ll(\Bb)$ is a disjoint union of $0$-priority regular languages, i.e. $\Gamma$ is a disjoint union of $\Gamma_1,\dots,\Gamma_k$ ($k \ge 1$) such that 
\begin{itemize}
\item for each $u \in \Sigma^\ast$, $\Aa$ outputs a word in $\Gamma_1^\ast \cup \dots \Gamma_k^\ast$,

\item $\Ll(\Bb)$ is a union of languages $L_1,\dots,L_k$ satisfying that $L_i \subseteq (\Gamma_i \times \{0,1\})^\ast$ is a $0$-priority regular language for each $i$.
\end{itemize}

For each $i$, let $\Gamma_i$ be ordered as $\gamma_{i,1}\dots \gamma_{i,l_i}$ under which $L_i$ is a $0$-priority regular language.

For each $i$, suppose $\Bb_i$ is a $0$-priority finite automaton accepting $L_i$ and $Acyc_{i,j}$($1 \le j \le l_i+1$) is the set of $0$-cyclic and $(\gamma_{i,j},0)$-acyclic states in $\Bb_i$.

Then from  the PCA $\Dd$, a PMA $\Cc$ can be constructed such that the counters of $\Cc$ correspond to the set of $0$-cyclic states in all these $\Bb_i$'s, and these counters are ordered as follows,
\begin{center}
$
\begin{array}{c}
   Acyc_{1,1}(Acyc_{1,2} \setminus Acyc_{1,1}) \dots (Acyc_{1,l_1} \setminus Acyc_{1,l_1-1}) Acyc_{1,l_1+1} \dots \\
Acyc_{k,1}(Acyc_{k,2} \setminus Acyc_{k,1}) \dots (Acyc_{k,l_k} \setminus Acyc_{k,l_k-1}) Acyc_{k,l_k+1}.
\end{array}$
\end{center}

In the PCA $\Dd$, after the transducer $\Aa$ nondeterministically chooses an index $i$ and outputs a string in $\Gamma^\ast_i$, only the $0$-priority finite automaton $\Bb_i$ is used and the other automata $\Bb_j$ for $j \ne i$ remain idle, thus the values of the counters before $Acyc_{i,1}$ in the above ordering are always zero, and the updates of the counter values corresponding to the states $Acyc_{i,1},\dots, Acyc_{i,l_i} \setminus Acyc_{l_i-1} Acyc_{l_i+1}$ can still be fulfilled using the restricted zero tests of PMAs. 

%\vspace*{-2mm}
\section{Application to the analysis of array-accessing programs}
\vspace*{-2mm}

In this section, we demonstrate how to apply class automata with priority class condition to the algorithmic analysis of array-processing programs considered in \cite{ACW09}. The notations of this section follow those in \cite{ACW09}.

An array $A$ is a list $(A[1].s, A[1].d)\dots (A[n].s, A[n].d)$ such that $A[i].s \in \Sigma$ and $A[i].d \in \mathds{D}$ for each $i: 1 \le i \le n$.

The syntax of array-accessing programs over an array $A$ are defined by the following rules\footnote{The nondeterministic-choice rule $if \ \ast \ then\ P \ else\ P$ is not included here for simplicity}:
\begin{center}
$\begin{array}{c}
  P :: = skip \mid \{ P \} \mid b := B  \mid p := IE \mid v := DE   \mid \\
  {\rm if} \ B \ {\rm then} \ P \ {\rm else}\ P \mid {\rm for} \ i := 1 \ to \ length(A) \ {\rm do} \ P \mid P;P
\end{array}$
\end{center}

where
\begin{itemize}
\item $i,j,i_1,j_1,\dots$ are loop variables, $p,p_1,\dots$ are index variables, $v,v_1,\dots$ are data variables, and $b,b_1,\dots$ are Boolean variables,

\item $s,s_1,\dots \in \Sigma$ and $c,c_1,\dots \in \mathds{D}$ are constants,

\item $IE:: = p \mid i$ are index expressions, $SE::=s \mid A[IE].s$ are $\Sigma$-expressions, $DE::=v \mid c \mid A[IE].d$ are data expressions, and
$B$ are Boolean expressions defined by the following rules,
\begin{center}
$B::= true \mid false \mid b \mid B \ and \ B \mid not \ B \mid IE = IE \mid IE < IE \mid DE = DE \mid DE < DE \mid SE = SE$.
\end{center}
\end{itemize}

A \emph{state} of the array-accessing program $P$ is an assignment of values to the variables in $P$.

A \emph{Boolean state} of the program $P$ is an assignment of values to the Boolean variables in $P$.

The \emph{initial state} of the program $P$ is a state such that
\begin{itemize}
\item all the Boolean variables have value $false$;

\item all the loop and index variables have value $1$;

\item all the data variables have the value the same as the first element of $A$.
\end{itemize}

A \emph{loop-free} program is a program containing no loops, namely a program formed without using the rules ``${\rm for}\ i:=1 \ {\rm to}\ length(A) \ {\rm do}\ P$''.

The \emph{Boolean state reachability} problem is defined as follows: Given a program $P$ and a Boolean state $m$ of $P$, whether there is an array $A$ such that $m$ is reached from the initial state after the execution of $P$ over $A$.

\emph{Restricted $ND_2$ programs} are programs of the following form,

%\vspace*{-1mm}
\begin{verbatim}
    for i:=1 to length(A) do
    { 
       P1;
       for j:=1 to length(A) do
       { 
          if A[i].d=A[j].d then
                    P2
          else
                    P3
       };
       P4
    }
\end{verbatim}
%\vspace*{-1mm}
\noindent such that
\begin{itemize}
\item $P1,P2,P3,P4$ are loop-free,
\item $P1,P2,P3,P4$ do not use index or data variables,
\item $P1,P2,P3,P4$ do not refer to the order on indices or data.
\end{itemize}
 
\begin{theorem}[\cite{ACW09}]\label{thm-rest-nd-2}
The Boolean state reachability problem is decidable for Restricted $ND_2$ programs satisfying the following additional condition:
\begin{quote}
$P3$ does not refer to $A[j]$, i.e. it does not contain the occurrences of $A[j].s$ or $A[j].d$.  
\end{quote}
\end{theorem}

The idea of the proof of Theorem \ref{thm-rest-nd-2} is to reduce the Boolean state reachability problem to the nonemptiness of extended data automata $\Dd=(\Aa,\Bb)$ (c.f. Remark \ref{rem-da-eda-ca}) such that 
\begin{itemize}
\item $\Aa$ guesses an accepting run of the outer-loop of $P$ over an array $A$, 

\item $\Bb$ corresponds to the inner loop and verifies the consistency of the guessed run. 
\end{itemize}

Roughly speaking, $\Bb$ can be constructed from $P2$ and $P3$ such that
\begin{itemize}
\item $P2$ corresponds to the one-transitions in $\Bb$,
\item $P3$ corresponds to the zero-transitions in $\Bb$.
\end{itemize}

The restriction that $P_3$ does not refer to $A[j]$ in Theorem \ref{thm-rest-nd-2} is crucial, because in extended data automata, the labels are omitted in zero-transitions of the class condition $\Bb$. 

On the other hand, as we have shown, PCAs, i.e. class automata with priority class conditions, do not omit the labels in zero-transitions and strictly generalize extended data automata. So naturally, by using PCAs, we should be able to show that the Boolean state reachability problem is decidable for a larger class of programs than those in Theorem \ref{thm-rest-nd-2}.

Similar to the construction of extended data automata from Restricted-$ND_2$ programs satisfying the additional condition in Theorem \ref{thm-rest-nd-2}, we have the following result.

\begin{lemma}\label{lem-rest-nd-2-ca}
For a Restricted-$ND_2$ program $P$ and a Boolean state $m$, a class automaton $\Dd=(\Aa,\Bb)$ can be constructed such that $m$ is reached from the initial state after the run of $P$ over an array $A$ iff  the array (data word) $A$ is accepted by $\Dd$.
\end{lemma}

In principle, the Boolean reachability problem is decidable for Restricted-$ND_2$ programs $P$ satisfying the additional condition that the class automaton $\Dd=(\Aa,\Bb)$ constructed from $P$ in Lemma \ref{lem-rest-nd-2-ca} is a class automaton with priority class condition. However, this condition is in some sense a semantical condition, since the construction of the automaton $\Dd$ from $P$ has an exponential blow-up. In the following, we demonstrate how to define a simple syntactic condition for $P3$ which guarantees that $\Dd$ constructed from $P$ is a PCA.

The \emph{$0$-priority restricted-$ND_2$} program is a Restricted-$ND_2$ program satisfying the following condition:
%\vspace*{-1mm}
\begin{quote}
Either $P3$ does not refer to $A[j]$, i.e. it does not contain the occurrences of $A[j].s$ or $A[j].d$, or there are a set of constants $s1,\dots, sr \in \Sigma$ such that $P3$ is a program of the following form,
%\vspace*{-3mm}
\begin{verbatim}
  if BB then
    if A[j].s =s1 then
       PA1
    else if A[j].s=s2 then
       PA2
    ...
    else if A[j].s=sr then
       PAr
    else skip
 else skip
\end{verbatim}
%\vspace*{-2mm}
such that 
\begin{itemize}
\item $BB$ is a conjunction of literals, i.e. $b$ or $not \ b$ for Boolean variables $b$,
\item $PA1, PA2,\dots, PAr$ are compositions of the assignments $b:=true$ or $b:=false$ for Boolean variables $b$,
\item Each $PAi$ for $1 \le i \le r$ is \emph{nontrivial} in the sense that there is a Boolean variable $b$ such that either  $b$ is a conjunct of $BB$ and the assignment $b:=false$ is in $PAi$, or $not \ b$ is a conjunct of $BB$ and the assignment $b:=true$ occurs in $PAi$.
\end{itemize}
\end{quote}

%\vspace*{-2mm}
\begin{remark}
The $0$-priority restricted-$ND_2$ programs subsume the Restricted-$ND_2$ programs satisfying that $P3$ does not refer to $A[j]$. A slightly more general syntactic condition than the above can be defined, which we choose not to present here, since the condition is rather tedious, and we believe that the simple condition presented above already sheds some light on the usefulness of PCAs.
\end{remark}

\begin{example}
The following program to describe the property ``for any two occurrences of the letter $a$ with the same data value in $A$, there is an occurrence of the letter $b$ between them with a different data value'' (c.f. Example \ref{exm-pca}) is an example of $0$-priority restricted-$ND_2$ programs. Intuitively, 
\begin{itemize}
\item the Boolean state $b1=true, b2=false, b3=false$ corresponds to the state $q_0$ in Figure \ref{fig-pca-example}(a), the Boolean state $b1=false,b2=true,b3=false$ corresponds to the state $q_1$, and the Boolean state $b1=false,b2=false,b3=true$ correspond to the Boolean state $q_2$;
\item the outer loop selects a position $i$ and the inner loop verifies that the class string corresponding to the data value $A[i].d$ satisfies the class condition.
\end{itemize} 

%\vspace*{-2mm}
\begin{verbatim}
  for i:=1 to length(A) do 
  {                                              
    if not b3 then    %the sink state q2 is not reached yet
      b1: = true; b2:=false
    else
      skip
    for j:=1 to length(A) do
    { if A[i].d = A[j].d then
       { if A[j].s=a then
             if b1 and not b2 and not b3 then
                b1:=false; b2:=true
             else if not b1 and b2 and not b3 then
                b2:=false; b3:=true
             else skip
          else skip
       }
       else 
       { if not b1 and b2 and not b3 then
             if A[j].s = b then
                b2:=false; b1:= true
             else skip
         else skip
       }
    }
 }
\end{verbatim}
 %\vspace*{-2mm}
An array $A$ satisfies the property iff the Boolean state $b1 = true, b2=false, b3=false$ or the state $b1=false, b2=true, b3=false$ is reached from  the initial state after the run of the above program over the array $A$.
\end{example}

\begin{theorem}\label{thm-0-priority-prog}
The Boolean state reachability problem is decidable for $0$-priority restricted-$ND_2$ programs. 
\end{theorem}

\medskip

\noindent {\bf \large Acknowledgement}. The author thanks Anca Muscholl for introducing him to this field. The author also thanks Luc Segoufin, St\'{e}phane Demri, and Miko{\l}aj Bojan\'{c}zyk for the discussions and suggestions. Last but not the least, the author thanks anonymous referees for their valuable suggestions and comments.

%\nocite{*}
\vspace*{-2mm}
\bibliographystyle{eptcs}
\bibliography{data}
\end{document}